\documentclass{article}
\usepackage[utf8]{inputenc}
\usepackage{amsmath}
\usepackage{amsfonts}
\usepackage{amssymb}
\usepackage{multirow}
\usepackage{comment}
\usepackage{algorithm}
\usepackage{algpseudocode}
\usepackage{graphicx}

\newcommand{\bs}{\mathbf}
\newcommand{\E}{\mathbb{E}}

\title{An Empirical Bayes Jackknife Regression Framework for Covariance Matrix Estimation}
\author{Huiqin Xin, Sihai Dave Zhao}
\date{}

\begin{document}

\maketitle

\section{Introduction}
Estimating the covariance matrix is a fundamental statistical problem with broad applications across various fields. For instance, in portfolio management, accurately estimating covariances among assets is crucial for risk assessment and optimization. In genomics, covariance matrix estimation plays a key role in constructing gene networks. However, when the number of features is comparable to or even exceeds the sample size, the traditional sample covariance matrix often exhibits performance degradation. As a result, high-dimensional covariance matrix estimation has emerged as a significant challenge, leading to the development of various advanced estimation methods to address this issue.

Among these methods, the recently proposed compound decision approach \cite{xin2020nonparametric} has demonstrated strong performance in both statistical simulations and real datasets. This method frames covariance matrix estimation as a compound decision problem, aiming to determine an optimal decision rule for a large number of parameters simultaneously. By introducing a new class of decision rules for covariance estimation and leveraging nonparametric empirical Bayes g-modeling to approximate the optimal rule, this approach offers distinct advantages over existing methods.

First, unlike thresholding \cite{cai2011adaptive}, banding \cite{li2017estimation, liu2020}, and low-rank models \cite{fan2019robust}, it does not impose specific structural assumptions on the population covariance matrix. Second, in contrast to commonly used rotation-invariant estimators \cite{ledoit2020quadratic, lam2016nonparametric}, which retain sample eigenvectors while adjusting eigenvalues, this approach mitigates the issue of sample eigenvectors deviating significantly from true eigenvectors in high-dimensional settings \cite{mestre2008on}. Numerical results further support that, in some cases, the compound decision approach achieves more precise covariance estimation.

However, the nonparametric g-modeling approach has certain limitations. It assumes that the data follows a Gaussian distribution, which restricts its applicability in scenarios where the data is non-normally distributed or the likelihood is unknown. Additionally, numerical results indicate that its performance is not always the most competitive in certain cases. Another drawback of matrix shrinkage via the g-modeling method is its computational cost. As the sample size increases, the complexity of the data likelihoods grows, requiring higher computational precision. Moreover, the memory cost scales cubically with the number of features, posing a significant challenge for large-scale applications.

In this paper, we introduce a new empirical Bayes framework for covariance estimation that overcomes these limitations. In the context of mean estimation, where data replications are generated from an unknown likelihood, the optimal Bayes decision rule can be approximated using regression algorithms by treating one observation as the response and the remaining ordered data replicates as features \cite{ignatiadis2021empirical}.

Building on this idea, our approach applies this technique to covariance matrix estimation. We first partition the samples to construct data replicates for each covariance parameter. Then, machine learning regression algorithms are used to approximate the optimal Bayes rule by leveraging these constructed features and responses. Since this method treats covariance estimation as a regression problem, it does not require prior knowledge of the data distribution. Numerical results demonstrate that our approach outperforms the g-modeling shrinkage method, achieving higher computational efficiency.

\section{Approach}
\subsection{Jackknife regression}
Suppose we have $n$ $p-$dimensional data samples $\bs{X}_1,\ldots, \bs{X}_n$. Each $\bs{X}_i$ is independently generated from a distribution with mean of zeros and covariance matrix $\bs{\Sigma}$. Our goal is to find an estimator $\bs{\delta}(\bs{X})$ of $\bs{\Sigma}$, minimizing the Frobenius risk
\begin{equation}\label{frobenius}
    R(\bs{\delta},\bs{\Sigma}) = \E [\sum_{j,k=1}^p(\delta_{jk}(\bs{X})-\sigma_{jk})^2]
\end{equation}
where $\sigma_{jk}$ represents $(j,k)$-th entry of $\bs{\Sigma}$, $j,k=1,\ldots,p$.

Compound decision theory solves the problem of simultaneously estimating a sequence of parameters. In \cite{xin2020nonparametric}, covariance matrix estimation is treated as a compound decision problem based on the fact that, the problem of minimizing the Frobenius risk \eqref{frobenius} is equivalent to minimizing the squared loss of its vector estimator $(\delta_{11}(\bs{X}),\ldots, \delta_{pp}(\bs{X}))$ for parameter vector $(\sigma_{11},\ldots,\sigma_{pp})$. 

Then \cite{xin2020nonparametric} generalizes the class of separable decision rules, which are commonly used in compound decision theory \cite{jiang2009general}, to covariance estimation. In compound decision problem where data $\bs{Y}=(Y_1,\ldots, Y_n)$ are generated from their means $\bs{\theta}=(\theta_1,\ldots,\theta_n)$, separable rule $\delta_i(Y)=t(Y_i)$ means the function on data relevant to each parameter. Applying it on covariance estimation, with given data $Z_1,\ldots,Z_n$, the generalized separable rule on off-diagonal and on-diagonal entries of covariance matrix is 
\begin{equation}\label{decision_rule}
    S=\{\bs{\delta}: \delta_{kj}=\delta_{jk}=t_{od}(Z_{\cdot j},Z_{\cdot k}),1\leq k<j\leq p. \quad \delta_{jj}=t_d(Z_{\cdot j}),j=1,\ldots,p\}
\end{equation}

Suppose $\sigma_j$ represent true standard deviation of $j$-th feature, $\rho_{jk}$ is the correlation between $j$-th and $k$-th feature. $G_{od}$ and $G_d$ are the empirical distributions of $(\sigma_j,\sigma_k,r_{jk})$, $1\leq j<k\leq p$ and $\sigma_j$, $j=1,\ldots,p$. For each data $\bs Z_i$, $f_2(\cdot|\sigma_j,\sigma_k,r_{jk})$ is the density of each $(j,k)$-th pair of data $(Z_{ij},Z_{ik})$ with row standard deviation $\sigma_j$, column standard deviation $\sigma_k$ and correlation $r_{jk}$, $f_1(\cdot |\sigma_j)$ is the density of $Z_{ij}$ with standard deviation $\sigma_j$. \cite{xin2020nonparametric} shows that, by fundamental theorem of decision theory \cite{robbins1951asymp}\cite{zhang2003compound}, the optimal decision rules among the class of separable rules \eqref{decision_rule} minimizing \eqref{frobenius} are the following Bayes rules
\begin{align}\label{bayes}
    t_{od}^* &= \E_{(\sigma_j,\sigma_k,r_{jk}) \sim G_{od},(Z_{\cdot j},Z_{\cdot k})\sim \prod f_2}[(\sigma_j\sigma_k r_{jk})|Z_{\cdot j},Z_{\cdot k}] \\
    \quad t_{d}^* &= \E_{\sigma_j \sim G_{d},Z_{\cdot j}\sim \prod f_1}[\sigma_j^2|Z_{\cdot j}]
\end{align}
where $\prod f_2$ means the product of all pairs of $f_2(Z_{ij},Z_{ik}|\sigma_j,\sigma_k,\rho_{jk})$, $\prod f_1$ means the product of $f_1(Z_{ij}|\sigma_j)$. This optimal Bayes rule is based on the Bayes model where the parameters $(\sigma_j,\sigma_k,\rho_{jk})$ and $\sigma_j$ are randomly generated from "priors" $G_{od}$, $G_d$. Although they could be deterministic parameters, in compound decision theory, it is useful to establish the result \eqref{bayes} on this Bayes model \cite{jiang2009general}.

In previous literature \cite{xin2020nonparametric}, with unknown likelihood functions $f_1$,$f_2$, the oracle optimal Bayes rules \eqref{bayes} are approximated by replacing unknown prior with estimated prior distributions $\hat G_{od}$, $\hat G_d$. Nevertheless, in many situations, the distribution of data is unknown and the existing $g-$modeling approach is misspecified. 

Unavailability of $f_1, f_2$ makes approximating the functions $t_{od}^*(Z_{\cdot j}, Z_{\cdot k})$, $t_d^*(Z_{\cdot j})$ difficult. Fortunately, in mean parameter estimation problem, \cite{ignatiadis2021empirical} shows that when there exists replicated data observations, the posterior mean of parameters conditioning on data is equivalent to the posterior mean of another independently generated data point conditioning on existing data. Based on this, without knowing data distribution, the optimal Bayes rule can be approximated by regression algorithms where the replicated data is manually split into features and response. 

This mechanism can be applied in covariance estimation. In our problem, suppose now we have $Z_1',\ldots, Z_n'$ that are independent from $Z_1,\ldots,Z_n$ following the same distribution. For any $1\leq j<k\leq p$, $s_{jk}'$ is the sample covariance of $(Z_{\cdot j}, Z_{\cdot k})$. Since $s_{jk}'$ has mean $\sigma_{jk}$, it can be shown that the conditional mean of $s_{jk}'$ given $(Z_{\cdot j},Z_{\cdot k})$ is equivalent to the optimal Bayes rule \eqref{bayes}.  
\begin{align*}
        &\E [s_{jk}'|(Z_{\cdot j},Z_{\cdot k})]\\
    =&\E [\E[s_{jk}'|(\sigma_j,\sigma_k,\rho_{jk})]|(Z_{\cdot j},Z_{\cdot k})]\\
    =&\E [\sigma_j\sigma_k \rho_{jk}|(Z_{\cdot j},Z_{\cdot k})] \\
    =&t^*_{od}(Z_{\cdot j},Z_{\cdot k})
\end{align*}
Similarly, for variances, 
\begin{align*}
        &\E [s_{jj}'|Z_{\cdot j}]\\
    =&\E [\E[s_{jj}'|\sigma_j]|Z_{\cdot j}]\\
    =&\E [\sigma_j^2|Z_{\cdot j}] \\
    =&t^*_{d}(Z_{\cdot j})
\end{align*}
Therefore, \eqref{bayes} could be approximated by regressing $s_{jk}'$ on $(Z_{\cdot j},Z_{\cdot k})$ and regressing $s_{jj}'$ on $Z_{\cdot j}$. 

This implies that we could use data replicates to construct regression problem and estimate true covariances by in-sample prediction. In our case, data replicates could be constructed by manually dividing all $n$ data samples into $M$ groups, $\bs{X}^{(m)}$, $m=1,\ldots,M$, each has $n/M$ samples with identical and independent distribution. For each $m$, we construct data points as follows
\begin{equation}\label{data1}
    ((X_{\cdot 1}^{(-m)},X_{\cdot 2}^{(-m)}), s_{12}^{(m)}),\ldots,((X_{\cdot p-1}^{(-m)},X_{\cdot p}^{(-m)}), s_{p-1,p}^{(m)})
\end{equation}
for off-diagonal covariances, and
\begin{equation}\label{data2}
    (X_{\cdot 1}^{(-m)}, s_{11}^{(m)}),\ldots,X_{\cdot p}^{(-m)}, s_{pp}^{(m)})
\end{equation}
for on-diagonal variances. $X_{\cdot j}^{(-m)}=\{X_{\cdot j}^{(l)}\}_{l\neq m}$ represents the j-th feature for all data groups excluding $m$-th group.

For the above data \eqref{data1}\eqref{data2}, $(X_{\cdot j}^{(-m)},X_{\cdot k}^{(-m)})$ has $2n(M-1)/M\sim O(n)$ features and $(X_{\cdot j}^{(-m)})$ has $n(M-1)/M\sim O(n)$ features. The number of features is comparable with sample sizes $p(p-1)/2$ and $p$. Doing regression on these high-dimensional data directly will result in high prediction error by curse of dimensionality in machine learning \cite{hastie2009elements}. Therefore, it is beneficial to reduce the number of features in our algorithm. 

It is known that for Gaussian distribution, $\bs\eta_{jk}^{(l)}=(s_{jj}^{(l)},s_{kk}^{(l)},s_{jk}^{(l)})$ is the sufficient statistics of $(\bs X_{\cdot j}^{(l)},\bs X_{\cdot k}^{(l)})$ with respect to the parameter $(\sigma_j,\sigma_k,\rho_{jk})$. In this case, it is feasible to use $\bs{\eta}_{jk}^{(l)}$ as features for Gaussian distribution without losing information. Even without Gaussian distribution assumption, we can use $\bs{\eta}_{jk}$ to predict covariance $\sigma_{jk}$. Thus the data with reduced dimension becomes  
\begin{equation}
    (\bs{\eta}_{12}^{(-m)}, s_{12}^{(m)}),\ldots,(\bs{\eta}_{p-1,p}^{(-m)}, s_{p-1,p}^{(m)})
\end{equation}
for off-diagonal covariances, where $\bs{\eta}_{jk}^{(-m)} = \{\bs{\eta}_{jk}^{(l)}\}_{l\neq m}$ represent the $3(M-1)$-dimensional features constructed from remaining data groups other than $m$-th group, and
\begin{equation}
    (s_{11}^{(-m)}, s_{11}^{(m)}),\ldots,(s_{pp}^{(-m)}, s_{pp}^{(m)})
\end{equation}
for on-diagonal variances, where $\bs{s}_{jj}^{(-m)}=\{s_{jj}^{(l)}\}_{l\neq m}$ represent $(M-1)$-dimensional feature constructed from the remaining group other than $m$-th group.

With these data points, we could apply specific regression algorithms on them to derive a function $\hat g_{od}(\bs{\eta}_{jk}^{(-m)})$ which approximates Bayes rule $t_{od}^{*}$ and $\hat g_{d}(s_{jj}^{(-m)})$ to approximate Bayes rule $t_{d}^{*}$. Then the covariances are estimated by in-sample prediction $\hat \sigma_{m,jk} = \hat g_{od}(\bs{\eta}_{jk}^{(-m)})$ and $\hat \sigma_{m,jj} = \hat g_{d}(s_{jj}^{(-m)})$. 

The estimation for each data split $\hat \sigma_{jk}$ is the average of $\hat \sigma_{m,jk}$ for all $m=1,\ldots,M$. To reduce randomness, this data split procedure is repeated several times and the final covariance estimation is the averaged estimated $\hat\sigma_{jk}$. This technique of splitting data and computing the average is also applied in the nonparametric eigenvalue-regularized estimator where data is split to estimate eigenvectors and eigenvalues separately \cite{lam2016nonparametric}. The averaged matrix estimation does not guaranteed to be positive definite. So we make the positive definiteness correction as in \cite{xin2020nonparametric} at the end to get the final estimator. The whole estimation procedure is displayed in \eqref{alg1}. 

\begin{algorithm}
\caption{Estimate covariance by jackknife regression}\label{alg1}
\begin{algorithmic}
\Require $X_1,\ldots, X_n$
\For{$t=1,\ldots,T$}
\State Split all data into $M$ groups $\{X^{(m)}\}_{m=1}^M$.
\For{$m=1,\ldots,M$}
\State
\begin{enumerate}
    \item Construct features for $X^{(-m)}$ and calculate $\bs{S}^{(m)}$, the sample covariance matrix of $\bs{X}^{(m)}$.
    \item Approximate Bayes rule \eqref{bayes} by machine learning algorithm and get the fitted models $\hat g_{od}^{m}$ and $\hat g_{d}^{m}$.
    \item Calculate the in-sample prediction for covariances $\hat \sigma_{m,jk}^{t} = \hat g_{od}^{m}(\bs\eta_{jk}^{(-m)})$ and variances $\hat{\sigma}_{m,jj}^{t} = \hat g_{d}^{m}(s_{jj}^{(-m)})$.
\end{enumerate}
\EndFor
\State Average across all $m$ to get the estimation $\hat \sigma_{jk}^{t}=\sum_{m=1}^M \hat \sigma_{m,jk}^{t}$.
\EndFor
\State Average across all data splits to get the final estimation $\hat \sigma_{jk} = \sum_{t=1}^T\hat \sigma_{jk}^{t}$. 
\State Project the estimated covariance matrix onto positive definite space $P(\bs{\hat\Sigma})$. 
\end{algorithmic}
\end{algorithm}

One important step in \eqref{alg1} is to choose a regression algorithm for getting the approximated Bayes decision rule. As a supervised learning problem, there exist abundant regression models in machine learning can be applied here. Among these methods, we apply three powerful regression algorithms, kNN, clustered linear regression and decision tree in this paper. 

One efficient regression model we adopt in our problem is the so called clustered linear regression \cite{ari2002clustered}, which partitions all data into clusters and do linear regression separately in each cluster. It can be more accurate than classic linear regression. We have shown in \cite{xin2020nonparametric} that linear regression could be applied in covariance matrix and the estimation is similar to linear shrinkage method \cite{ledoit2004well}, which is the combination of sample covariance matrix and identity matrix. The linear coefficients are computed by data and used on all entries of the matrices. This method is easy to compute and is shown to be more accurate than sample covariance matrix. However, global linear coefficients may not be accurate enough because the intensity of shrinkage is same for all covariances. In clustered linear regression, local linear coefficients are determined for each cluster and can be seen as the estimated first order partial derivatives with Taylor expansion of the function \eqref{bayes}. In our jackknife regression framework, as an extension of linear regression, we apply clustered linear regression on covariance matrix estimation and estimate the local linear coefficients on constructed features $\bs{\eta}_{jk}$ and $s_{jj}$.  

The first step of clustered linear regression is do clustering on all points, with a given the number of clusters $K_c$. With more clusters, the localization is more subtle but there are less data points in each cluster. With less cluster, the linear coefficients are rough but less likely to be overfitted. So it is necessary to choose the appropriate $K_c$. In cluster analysis, "elbow method" is the common measure to determine the optimal number of clusters. The elbow point means as $K_c$ grows, the within cluster sum of squared errors (WSS) decreases, and there exists some point that the decrease of WSS starts to diminish and the $K_c$ v.s. WSS curve becomes flat. This "elbow" point is determined to be the optimal $K_c$.

Another simple but powerful machine learning regressor is kNN. With the number of nearest points $k$, the estimation $\hat\sigma_{m,jk}^t$ is the averaged $s_{j'k'}^{(m)}$ for the $k$ nearest points $\bs{\eta}_{j'k'}^{(-m)}$ from $\bs{\eta}_{jk}^{(-m)}$. Here we measure the distance between any two points $\bs{\eta}_{j_1k_1}, \bs{\eta}_{j_2k_2}$ by their Euclidean distance after each dimension of feature is scaled to be centered and have variance 1. kNN is related with MSG in some way. Its estimation $\hat\sigma_{jk}=\frac{1}{k}\sum_{(j'k')\in D_{jk}} s_{j'k'}^{(m)}$ is an average of other sample values, where the weight $\frac{1}{k}\mathbb{1}((j'k')\in  D_{jk})$ of each sample $(j',k')$ can be seen as a relaxation of its posterior weight used in MSG, as equation 10 \cite{xin2020nonparametric} shows.

The parameter $k$ can be determined by cross validation. In the beginning, after constructing features and responses by randomly split data, we get $p(p-1)/2$ data points for covariances and $p$ points for variances. For both covariance and variance models, all the data samples are randomly divided into 80 percent of training samples and 20 percent of testing samples. Different models are fitted on training samples with a sequence of $k$ and take the optimal $k$ which has the lowest sum of squared loss between fitted covariance $\hat\sigma_{jk}$ and sample covariance in test samples. 


\section{Simulation}
In this section, we run our jackknife regression framework combining with clustered linear regression, kNN and decision tree models, which are written as {\bf Clustered LR, KNN, Tree} in this section. In these regression methods, we adopt default parameter values in the simulation in order to improve computational efficiency. For clustered linear regression, number of clusters is set as 3 for variance models and 10 for covariance models. For kNN, the number of nearest points is set as $N^{1/2}$, $N$ is the number of data points in regression model. in tree model, the parameters are the default values in R `rpart` package. However, these parameters might not be the optimal choice which can be chosen by cross-validation. In our numerical experiments, the improvement is very limited. See more details in the support information. 

We also run other covariance matrix estimators as presented in \cite{xin2020nonparametric} to make comparison. These methods include seven competitors

{\bf MSGCor} The nonparametric empirical Bayes g-modeling approach with positive definiteness correction \cite{xin2020nonparametric}.

{\bf CorShrink} An empirical Bayes method aiming to estimate correlation matrix. We apply this method to estimate correlation matrix and multiply it with sample standard deviations \cite{dey2018corshrink}.   

{\bf Linear} Linear shrinkage estimator combining sample covariance matrix and identity matrix \cite{ledoit2004well}.

{\bf Adap} Adaptive thresholding estimator targeting to estimate sparse covariance matrix \cite{cai2011adaptive}.

{\bf QIS} Nonlinear shrinkage estimator which keeps eigenvectors and shrinks eigenvalues \cite{ledoit2020quadratic}.

{\bf NERCOME} Nonlinear shrinkage estimator splitting data to estimate eigenvalues and eigenvectors \cite{lam2016nonparametric}.

{\bf Sample} The sample covariance matrix.

as well two oracle estimators that are unobtainable because their computation contains unknown parameter in $\bs{\Sigma}$

{\bf OracNonlin} Optimal rotation-invariant estimator. 

{\bf OracMSG} Same as MSGCor except the sample grid points are consist of true unknown parameters $(\sigma_j,\sigma_k,\rho_{jk})$.

\begin{figure}
\begin{center}
\centerline{ \includegraphics[width=0.85\textwidth]{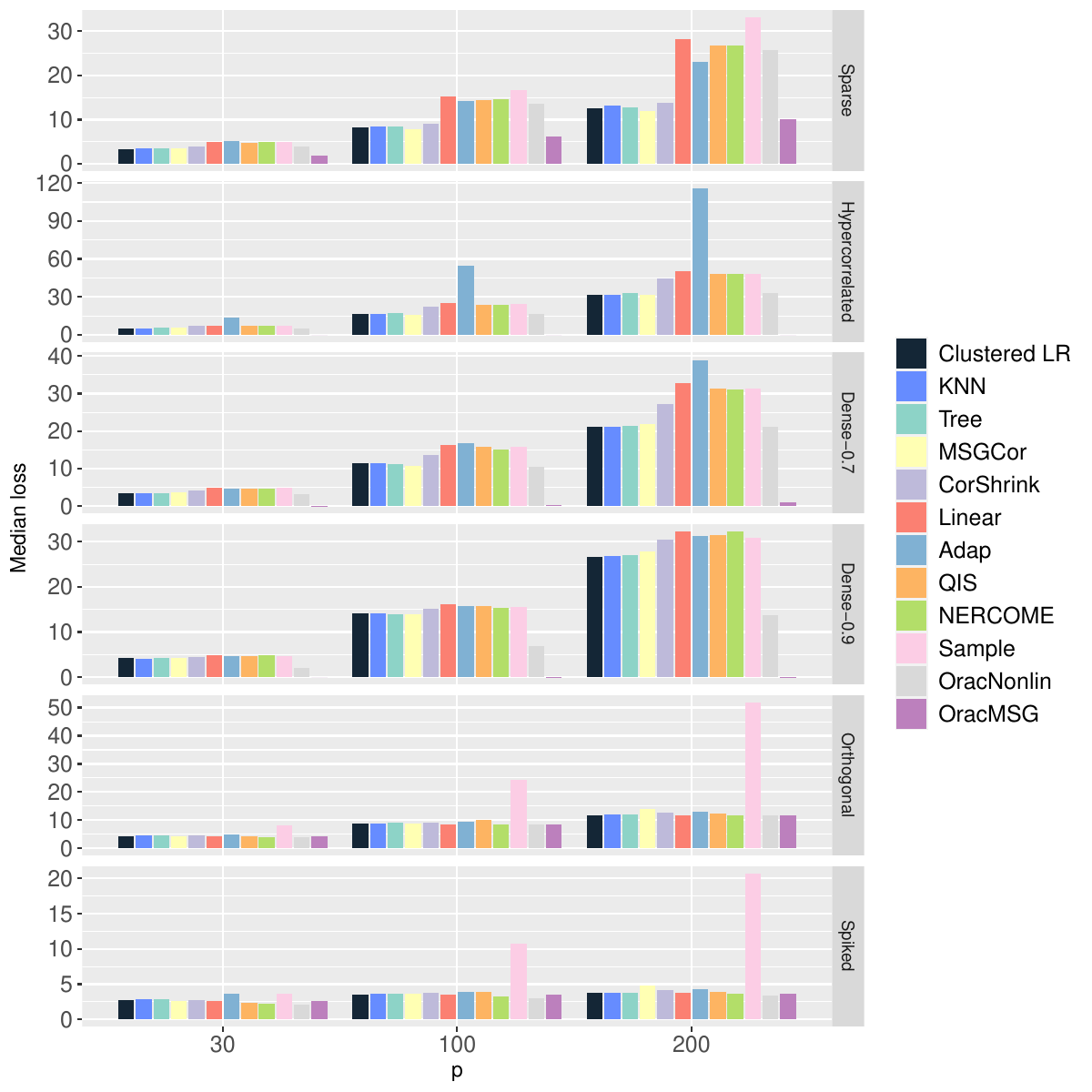}}
\end{center}
\caption{Median Frobenius norm errors over 200 replications for Gaussian distributed data. Sparse: Model 1; Hypercorrelated: Model 2; Dense-0.7: Model 3; Dense-0.9: Model 4; Orthogonal: Model 5; Spiked: Model 6.}
\label{fig:err_normal}
\end{figure}

Six different population covariance matrices are designed as in \cite{xin2020nonparametric}. For each design, data dimension is taken as $p=30,100,200$ and sample size $n$ is always 100. We first generate data from multivariate normal distribution. The median Frobenius loss across 200 replicates of data for all estimators is shown in \ref{fig:err_normal}.

The result \ref{fig:err_normal} shows that our approach has competitive performance in every matrix model. Among these three regression algorithms, clustered linear regression has slightly smaller error than the other two regression methods, but they has closed overall performance. For first four matrix models, Sparse, Hypercorrelated, Dense-0.7, Dense-0.9, MSGCor and CorShrink have comparable behavior than our jackknife regression approach, sometimes MSGCor can even beat our method. However, in the last two matrix models, Orthogonal and Spiked, our methods has obvious improvement comparing to MSGCor. Comparing with other estimators, Our method has dramatically lower error in first four models and nearly the lowest error in last two models.

\begin{figure}
\begin{center}
\centerline{ \includegraphics[width=0.85\textwidth]{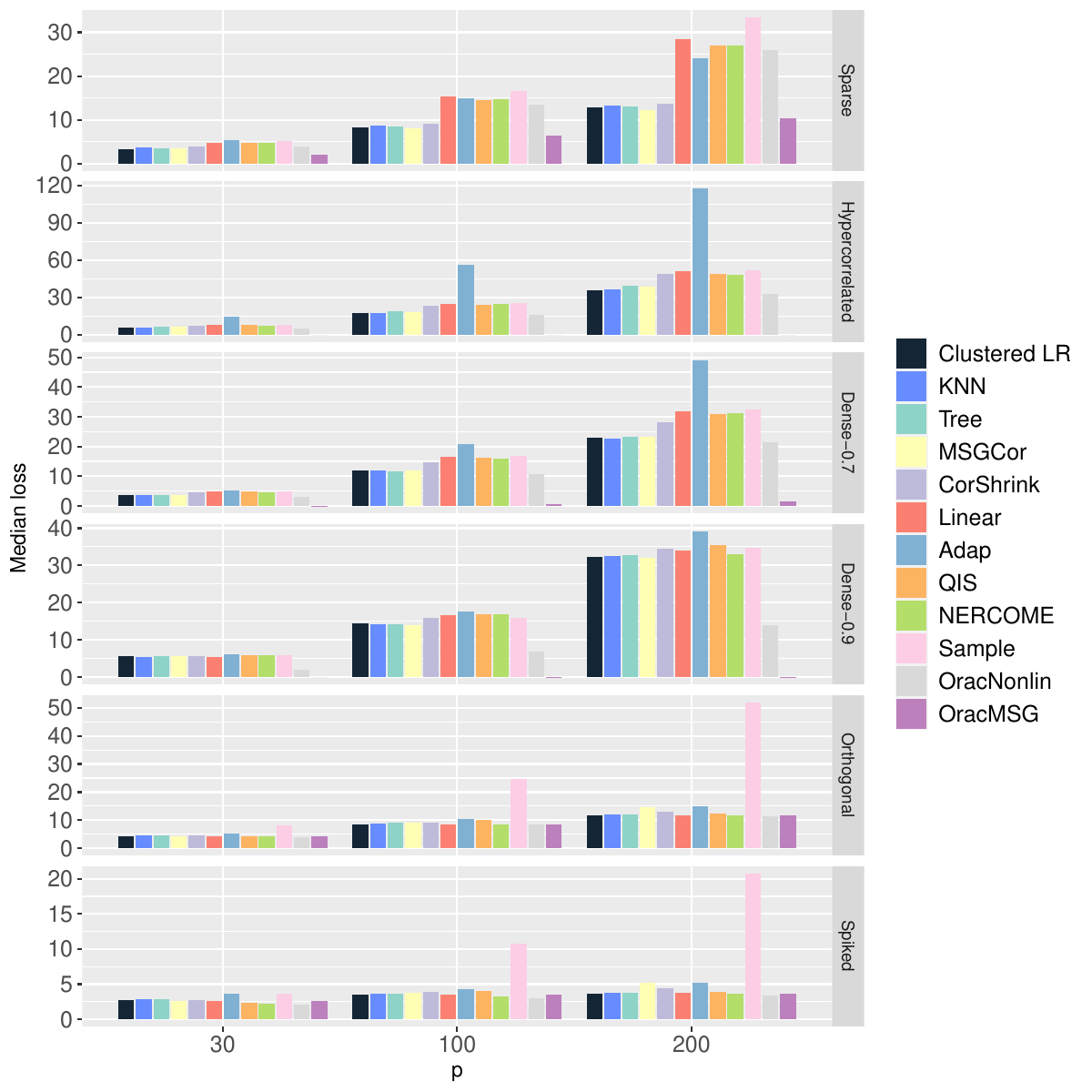}}
\end{center}
\caption{Median Frobenius norm errors over 200 replications for negative binomial data. Sparse: Model 1; Hypercorrelated: Model 2; Dense-0.7: Model 3; Dense-0.9: Model 4; Orthogonal: Model 5; Spiked: Model 6.}
\label{fig:err_nb}
\end{figure}

\begin{figure}
\begin{center}
\centerline{ \includegraphics[width=0.85\textwidth]{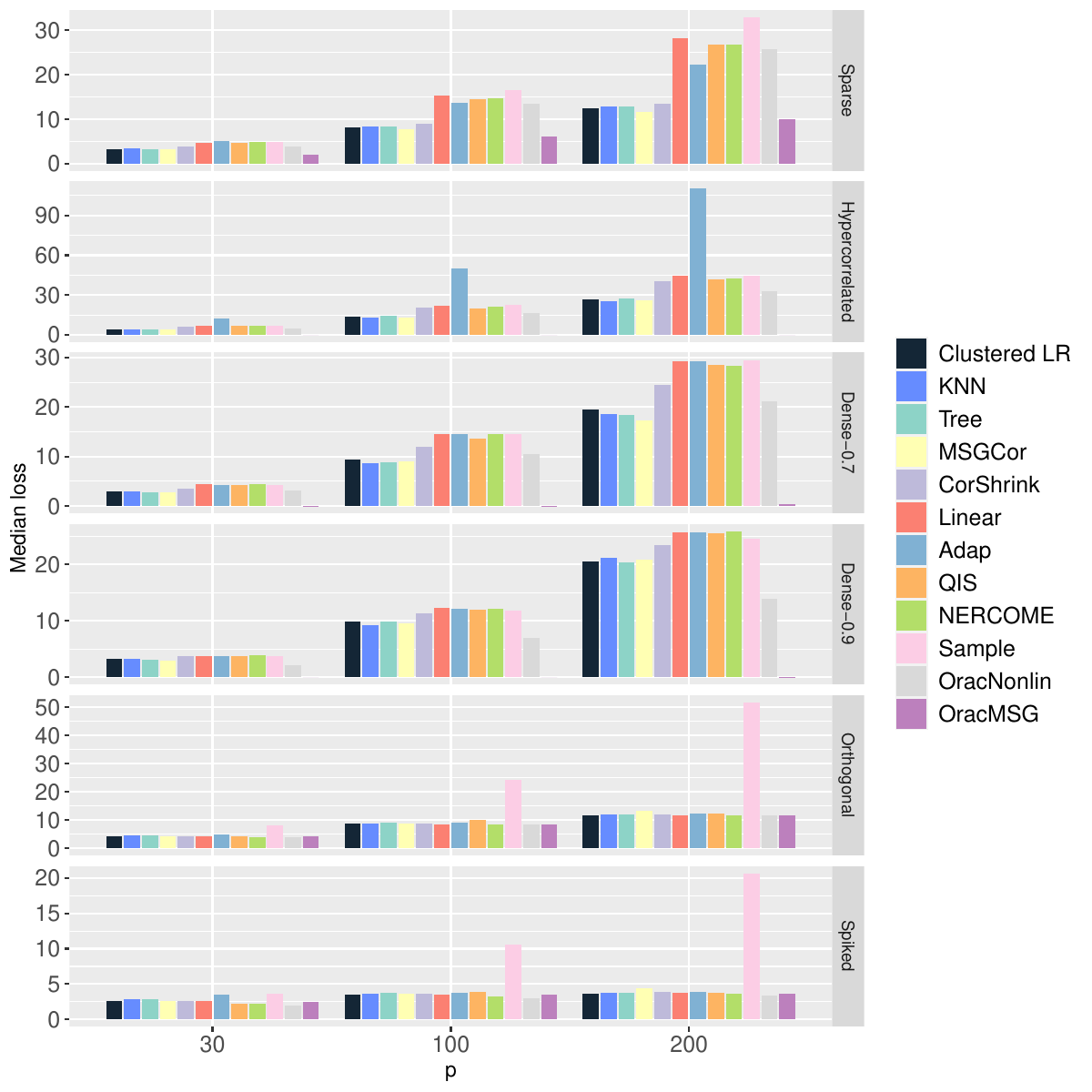}}
\end{center}
\caption{Median Frobenius norm errors over 200 replications for uniform data. Sparse: Model 1; Hypercorrelated: Model 2; Dense-0.7: Model 3; Dense-0.9: Model 4; Orthogonal: Model 5; Spiked: Model 6.}
\label{fig:err_unif}
\end{figure}

Our framework is not based on the assumption of normally distributed data. So we also interested in its behavior on non-Gaussian data. In the non-Gaussian simulation, data is generated as $\bs{X}=\bs{L}\bs{Y}$, where $\bs{L}$ is the Cholesky decomposition of the population covariance matrix. $\bs{Y}$ is the non-Gaussian $p$-dimensional sample where each entry is identically and independently generated from some univariate distribution. Here, the same as in \cite{xin2020nonparametric}, we consider two settings, negative binomial distribution with size 10 and mean 4, standard uniform distribution. The results are displayed as in \ref{fig:err_nb} and \ref{fig:err_unif}.   

According to the result \ref{fig:err_nb}\ref{fig:err_unif}, our estimator still has competitive behavior in non Gaussian data. It is surprising that MSGCor still behaves well in the first four matrix models. This phenomenon might be explained by the fact that the sample covariance matrix of non-normal multivariate data can be approximated by Wishart distribution \cite{kollo1995approximating}. Although the simulation data disobey normal assumption in MSGCor, it also makes our features $\bs{\eta}_{jk}=(s_{jj},s_{kk},r_{jk})$ insufficient. Therefore, it is worth exploring to construct more related features. However, in the last two models our approach has more improvement comparing to MSGCor, especially in negative binomial case.
 
\section{Data analysis}
In this section, we apply our framework on gene network construction. We do analysis on the RNA-sequencing data from the experiment on three different brain regions of 5 mice, amygdala, frontal cortex and hypothalamus. For each brain region, RNA-sequencing data are collected on 3 time points for each mouse. In this case, there are 15 samples in each group, except for 2 samples are missing in hypothalamus group. More details about the context are introduced in \cite{xin2020nonparametric}. We adopt the same procedure as in \cite{xin2020nonparametric} to pick the top 200 genes that differentiate the most across three regions and transform gene values to log-counts per million mapped reads. Our goal is to estimate the covariance between each two genes.

As in \cite{xin2020nonparametric}, we first investigate the accuracy of covariance matrix estimation. We split 15 samples of amygdala and frontal cortex regions into 10 training samples to compute covariance matrix estimators, as well as 5 testing samples to calculate their sample covariance matrix. The accuracy is measured by the Frobenius norm of the difference matrix between estimated covariance matrix and sample covariance matrix. This split procedure is repeated for 200 times, the median Frobenius error and interquantile ranges are shown in \ref{tab:tab1}. 

\begin{table}
\begin{center}
\caption{\label{tab:tab1}Median gene expression covariance matrix estimation errors (25\% and 75\% quantiles in parentheses). Bold text highlights the smallest median errors in each column.}
\begin{tabular}{rll}
  \hline
  Brain region & Amygdala & Frontal cortex \\
  \hline
  Clustered-lr & 2.24(1.96,2.60) & 2.23(2.09,2.50) \\
  KNN & 2.25 (1.95, 2.64) & \textbf{2.22(2.02,2.51)} \\
  Tree & \textbf{2.23(1.96,2.64)} & 2.24(2.09,2.53) \\
  MSGCor & 2.26 (2.03, 2.59) & 2.27 (2.14, 2.50) \\
  Adap & 2.64 (2.20, 3.08) & 2.39 (2.16, 2.66) \\
  Linear & 2.30 (2.11, 2.58) & 2.30 (2.16, 2.52) \\
  QIS & 2.53 (2.09, 2.98) & 2.38 (2.17, 2.64) \\
  NERCOME & 2.37 (2.14, 2.68) & 2.25 (2.11, 2.51) \\
  CorShrink & 2.27 (2.05, 2.56) & 2.31 (2.18, 2.50) \\
  Sample & 2.61 (2.33, 2.85) & 2.75 (2.60, 2.89) \\
  \hline
\end{tabular}
\end{center}
\end{table}

From the table \ref{tab:tab1}, it is observed that the regression method has the best performance among all estimators, even slightly better than MSGCor. KNN and Tree has the lowest median Frobenius error for amygdala and frontal cortex. Clustered linear regression has the second lowest error for both regions. This shows that in different situations, different regression models are favored.

We are also interested in building gene networks using our estimator. The technique is the same as \cite{xin2020nonparametric}, adaptive thresholding estimator \cite{cai2011adaptive} is used to determine the sparsity of all estimated correlation matrices. For all other methods, estimated correlations are truncated such that the truncated matrix has the same sparsity as adaptive thresholding estimator. For any two genes, they are connected if they have non-zero correlation. We plot out gene networks for amygdala and frontal cortex regions. It can be observed that for amygdala region, our three estimators show similar pattern as other networks, except for Linear and NERCOME which look different. For frontal cortex region, the network built by Tree has denser pattern than others.  

\section{Discussion}
In this paper, we split data into equal-sized groups and make regression with sample covariances of data in each group. In fact, it is unnecessary to evenly divide all samples. There might exist better strategy to split data. For example, in \cite{lam2016nonparametric} where data are split for estimating eigenvectors and eigenvalues separately, the group size is chosen by cross-validation. 

Both our framework and MSGCor \cite{xin2020nonparametric} aim to approximate the empirical Bayes rule \eqref{bayes}, but conditioning on different data. MSGCor approximates the Bayes rule conditioning the whole data, while jackknife regression framework conditions on part of data for each single model \eqref{bayes}. This means the oracle Bayes risk is lower in MSGCor. Nevertheless, our framework shows better approximation efficiency in some models and we average across all model estimations to make full use of data. One possible reason is, unlike \cite{xin2020nonparametric} which uses pseudolikelihood to estimate the prior, ignoring the dependency between different entries of the sample covariance matrix, the dependency does not affect the approximation in our regression problem.   

\end{document}